\documentclass[10pt,twocolumn]{article} 

\usepackage[utf8]{inputenc} 
\usepackage[spanish]{babel}

\usepackage{geometry} 
\usepackage{graphicx} 
\usepackage{paralist} 
\usepackage{verbatim} 
\usepackage{subfig} 
\usepackage{lipsum}  
\usepackage{hyperref}
\usepackage[superscript]{cite}  

\usepackage[format=plain,
            labelfont=it,
            textfont=it]{caption}

\usepackage{fancyhdr} 
\usepackage{varwidth}
\usepackage{authblk}
\newcommand{\filiacion}[2]{\affil[#1]{\protect\begin{varwidth}[t]{\linewidth}\protect\centering \normalfont#2 \protect\end{varwidth}}}
\newcommand{\autor}[2]{\author[#1]{\bf #2}}
\newcommand{\corresponding}[2]{\author[#1]{\bf #2\thanks{}}}
\newcommand\cauthemail[1]{\footnotetext{#1}}
\newcommand{\fecha}[1]{\date{\vspace{-1ex}\small{#1}}}
\newcommand{\titulo}[2]{\title{\bf{\large{#1 \\ \vspace{1.5ex} #2 }}}}
\newcommand{\esresumen}[1]{\small{#1 \par}\vspace{1.5ex}}
\newcommand{\pclaves}[1]{\small{\emph{#1} \par}\vspace{1.5ex}}
\newcommand{\enresumen}[1]{\small{#1 \par}\vspace{1.5ex}}
\newcommand{\keywords}[1]{\small{\emph{#1} \par}\vspace{1.5ex}}

\usepackage{adjustbox}
\usepackage{graphicx}

\usepackage{sectsty}
\allsectionsfont{\fontsize{10}{12}\sffamily\bfseries\upshape} 

\usepackage{titlesec}
\titlespacing*{\section}{0pt}{1.5ex}{0.8ex}
\titlespacing*{\subsection}{0pt}{1.2ex}{0.6ex}
\usepackage[nottoc,notlof,notlot]{tocbibind} 
\usepackage[titles,subfigure]{tocloft} 

\usepackage{booktabs} 
\usepackage{array} 
\makeatletter
\newcommand{\thickhline}{%
    \noalign {\ifnum 0=`}\fi \hrule height 1.5pt
    \futurelet \reserved@a \@xhline
}
\newcolumntype{"}{@{\hskip\tabcolsep\vrule width 1pt\hskip\tabcolsep}}
\makeatother
\newcolumntype{L}[1]{>{\raggedright\let\newline\\\arraybackslash\hspace{0pt}}m{#1}}
\newcolumntype{C}[1]{>{\centering\let\newline\\\arraybackslash\hspace{0pt}}m{#1}}
\newcolumntype{R}[1]{>{\raggedleft\let\newline\\\arraybackslash\hspace{0pt}}m{#1}}

\makeatletter
\renewcommand\@biblabel[1]{#1.}
\makeatother

\titulo{ESTUDIO ARQUEOASTRONÓMICO DE LAS IGLESIAS HISTÓRICAS DE LA GOMERA}{ARCHAEOASTRONOMICAL STUDY OF THE HISTORIC CHURCHES OF LA GOMERA}

\corresponding{1}{A. Di Paolo} 
\autor{2}{A. Gangui}

\filiacion{1}{
Universidad de Buenos Aires, Facultad de Ciencias Exactas y Naturales, Departamento de Física. Buenos Aires, Argentina.} 
\filiacion{2}{
Universidad de Buenos Aires, Facultad de Ciencias Exactas y Naturales, Argentina. \par CONICET - Universidad de Buenos Aires, Instituto de Astronomía y Física del Espacio (IAFE), Argentina.}

\fecha{Recibido: 22/03/18; Aceptado: 10/08/18} 

\begin{document}

\renewcommand{\abstractname}{}
\twocolumn[
  \begin{@twocolumnfalse}
   \maketitle
    \begin{abstract}\vspace{-12ex}
\centering\begin{minipage}{\dimexpr\paperwidth-6cm}

\esresumen{
En este trabajo discutimos la importancia de estudiar la orientación de las iglesias cristianas antiguas, como un complemento a la investigación histórica y cultural de los templos. Presentamos
resultados preliminares del análisis de la orientación espacial precisa de 38 iglesias coloniales de la isla canaria de La Gomera. La muestra sugiere que, aunque varias iglesias tienen una orientación
canónica, que entra en el rango solar, una gran proporción de ellas sigue un patrón de orientaciones que es compatible con la abrupta orografía de la isla y que, por lo tanto, se aleja de las
prescripciones contenidas en los textos de los escritores cristianos tempranos.
}
\pclaves{Palabras Clave: orientación de iglesias, templos cristianos, Astronomía.} 

\enresumen{
In this paper we discuss the importance of studying the orientation of ancient Christian churches, as a complement to the historical and cultural research of the temples. We present preliminary
results of the analysis of the precise spatial orientation of 38 colonial churches located on the Canary Island of La Gomera (Spain). The sample suggests that, although several churches have a
canonical orientation within the solar range, a large proportion of them follow a pattern of orientations that is compatible with the steep orography of the island and, therefore, contrasts with the
prescriptions contained in the texts of the early Christian writers.
}

\end{minipage}
\vspace{4ex}
 \end{abstract}
  \end{@twocolumnfalse}
]

\thispagestyle{empty}

\setcounter{footnote}{1}
\cauthemail{adriandipaolo@gmail.com}  

\section{INTRODUCCIÓN}

\begin{figure}[ht]
\centering
\includegraphics[width=0.48\textwidth]{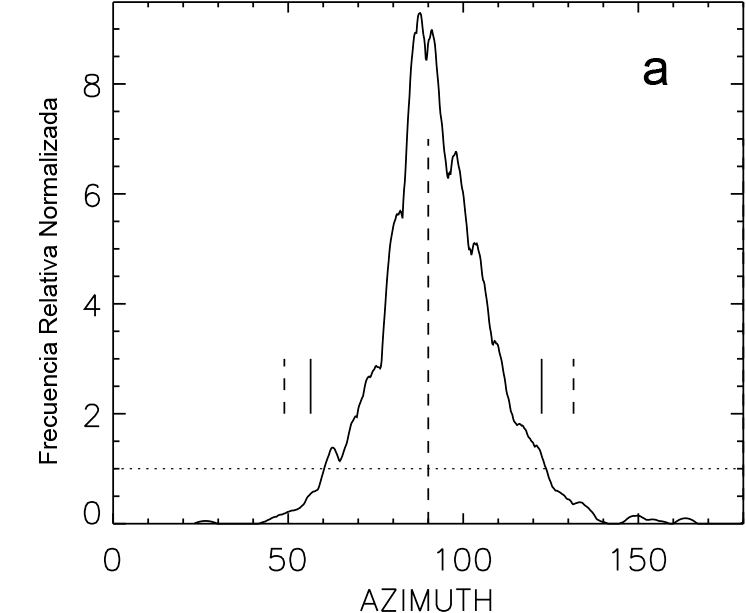}
\caption{Histograma de acimut de orientación de una muestra muy significativa de iglesias medievales europeas, que incluye las medidas de 1274 iglesias obtenidas de la literatura.\cite{gonzalez-15}
Nótese la concentración de orientaciones hacia el este correspondiente al equinoccio astronómico.} 
\label{fig:figura1}
\end{figure}

El estudio de las orientaciones de las iglesias medievales es, junto con las pirámides de Egipto, los megalitos europeos y las construcciones históricas de Mesoamérica, uno de los temas más antiguos
que se han trabajado en Arqueoastronomía. Trabajos recientes (por ejemplo, el de González-García\cite{gonzalez-15}) muestran que las prescripciones para la orientación hacia el oriente se siguieron de
forma muy sistemática en toda Europa, al menos durante la Edad Media (ver Fig. \ref{fig:figura1}).

Se sabe que la orientación espacial de las iglesias cristianas antiguas es una de las características más destacadas de su arquitectura. Excediendo los límites europeos, en muchos lugares remotos donde
llegaron los misioneros, hubo una marcada tendencia a orientar los altares de los templos en el rango solar. Es decir, el eje del templo, desde la puerta de entrada hacia el altar, se halla alineado
con los puntos en el horizonte por donde sale el Sol en diferentes días del año. Entre estos días, hay una marcada preferencia por los que corresponden a los equinoccios astronómicos, cuando los ejes
apuntan hacia el este geográfico.\cite{mccluskey-15, zimbron-15, gangui-etal-16}

Resulta interesante destacar que, salvo un número pequeño de trabajos dedicados a iglesias particulares, a sus orientaciones y a posibles eventos de iluminación, sobre todo en Inglaterra y en el
centro de Europa, no existen estudios sistemáticos sobre la orientación de los templos en períodos posteriores a la Edad Media, como el que nos ocupa, pues como veremos, la gran mayoría de las
iglesias y ermitas de la isla canaria de La Gomera se empezó a erigir décadas después de la conquista y colonización de la isla por los nobles al servicio de la corona de Castilla a partir del año
1440.\cite{diaz-05} En este trabajo presentamos un primer análisis de nuestras mediciones recientes de orientaciones, posiblemente astronómicas, de iglesias antiguas ubicadas en este pequeño y
relevante territorio insular. Nuestra intención es ver si en este territorio acotado, y ubicado lejos de la metrópoli, se respetaron los textos de los escritores y apologetas cristianos tempranos en
lo que respecta a la orientación {\it ad orientem} de la arquitectura sagrada.\cite{vogel-62}

\section{LAS ERMITAS E IGLESIAS DE LA GOMERA}

\begin{figure}[ht]
\centering

\includegraphics[width=0.15\textwidth]{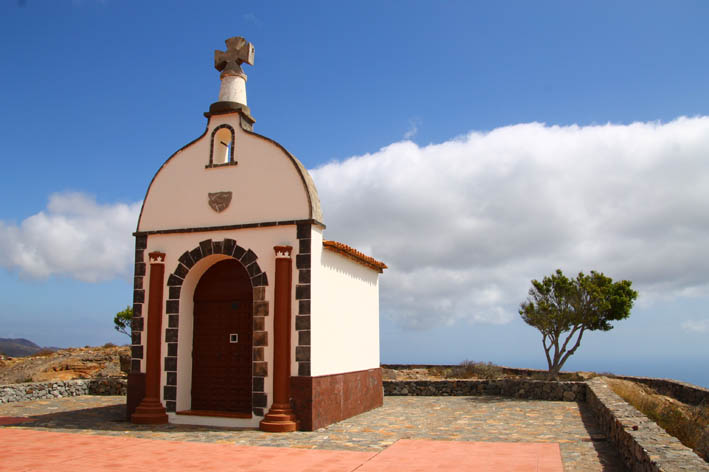}
\includegraphics[width=0.15\textwidth]{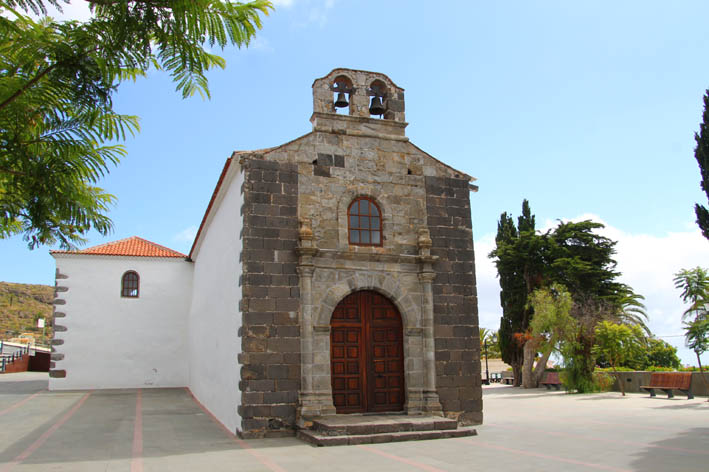}
\includegraphics[width=0.15\textwidth]{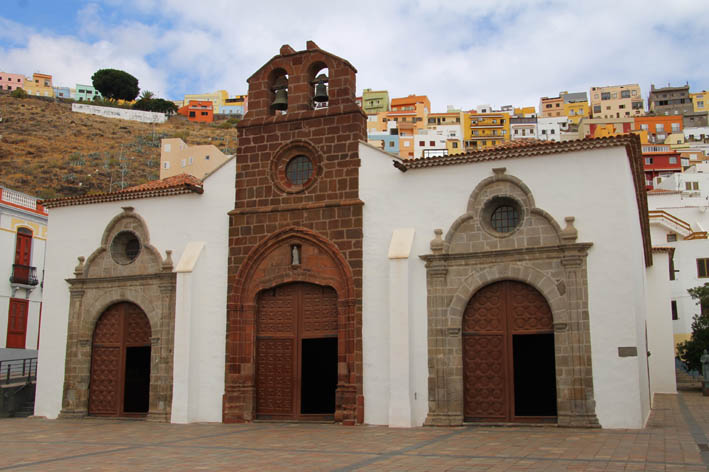}
\caption{Algunas iglesias características de la isla: a la izquierda San Isidro, emplazada en la cima de Roque Calvario, y al centro El Salvador, ambas ubicadas en el municipio de Alajeró.
  A la derecha Nuestra Señora de la Asunción, en San Sebastián, ciudad capital de La Gomera.} 
\label{fig:fig-iglesias}
\end{figure}

La arquitectura religiosa en la isla de La Gomera comenzó con la edificación de ermitas de factura sencilla y de recinto único, como la iglesia de San Isidro, en Roque Calvario (Alajeró), ubicada en
la cima de la Montaña de Tagaragunche, cercana de yacimientos arqueológicos indígenas (Fig. \ref{fig:fig-iglesias}). A algunas de ellas, con el paso del tiempo, se les fueron agregando capillas en la
cabecera, sacristías a sus lados y otros elementos de uso práctico. En general, estas construcciones no estuvieron sujetas a planes de ejecución estrictos y por ello, tanto su planta como su
estructura se levantaron de acuerdo a las necesidades del momento. Con el tiempo, algunas alcanzaron un cierto carácter monumental, con portadas en arco de medio punto, espadaña de uno o varios huecos
y techo a dos o cuatro aguas, la mayoría de las veces con tejas (Fig. \ref{fig:fig-iglesias}).

Dado el elevado número de estas construcciones históricas, cercano a los cuarenta y por lo tanto estadísticamente significativo, tenida en cuenta la superficie del territorio -de menos de 370 km$^2$-,
decidimos trabajar en La Gomera como un nuevo laboratorio de prueba donde estudiar la orientación de las iglesias canarias en los siglos inmediatamente posteriores a la conquista, como lo hicimos años
antes en la vecina isla de Lanzarote.\cite{gangui-etal-16} Nuestra motivación fue analizar si en la orientación de las iglesias en este territorio habían o no influido factores
como la presencia de la población aborigen, que tendría unos patrones de culto diferentes a los de los colonizadores recién llegados.\cite{belmonte-etal-94} 

La Tabla 1 muestra los datos obtenidos en una campaña de trabajo de campo. Se presentan las cantidades estándar (identificación de la iglesia o ermita y sus coordenadas) junto con su orientación, dada
por el acimut de los ejes de las construcciones medido en cada sitio, y luego corregido de acuerdo a la declinación magnética local, 
además de la altura angular del horizonte tomada a lo largo del eje de cada edificio en dirección al altar.
En aquellos sitios donde el horizonte estaba bloqueado (señalados con B en la última columna de la Tabla),
hicimos una reconstrucción del horizonte usando un modelo digital del terreno disponible en heywhatsthat.com. 
Nuestras mediciones se obtuvieron con brújulas de alta 
precisión. Los valores de la declinación magnética para distintos sitios de la isla oscilan entre 5$^{\circ}$27' y 5$^{\circ}$33' oeste. La precisión que tenemos con los acimut magnéticos medidos es
de 0,5$^{\circ}$ aproximadamente, por lo que la diferencia en declinación magnética entra bien dentro de nuestro error. Como una corroboración adicional, se verificaron las orientaciones medidas con
imágenes fotosatelitales, encontrándose pocas y mínimas divergencias.

\begin{table*}[t]  
\centering
\caption{Orientaciones de las iglesias de La Gomera, ordenadas por acimut creciente. Para cada construcción, la tabla muestra la ubicación, la identificación, la latitud y longitud
  geográficas (L y l), el acimut astronómico (a) y la altura angular del horizonte (h) tomados a lo largo del eje del edificio en dirección al altar (aproximados al 0.5$^{\circ}$ de error),
  con valores expresados en grados decimales. En la última columna, B señala los valores de h reconstruidos usando un modelo digital del terreno.}
\label{tabl:tab1}
\begin{adjustbox}{center}
\begin{tabular}{|llllrr|}
\hline
 Ubicación     & Nombre    &  L($^{\circ}$, N) &  l($^{\circ}$, O)      & a($^{\circ}$) & h($^{\circ}$)      \\ \thickhline
(1) El Cercado      & Virgen del Pino   &  28.11905        &  17.28493  &  2.5          &  2.5   \\ \hline
(2) Vallehermoso   & Virgen del Carmen &  28.15457       &   17.26933   &  2.5          & 13.5    \\ \hline
(3) Imada           & Santa Ana         &  28.08408        &  17.24084  &  2.5          & 19.0    \\ \hline
(4) Alajeró   & Ntra. Sra. del Buen Paso &  28.08695	&  17.24917     &  8.5        &  B 0.0   \\ \hline
(5) Las Rosas      &    Santa Rosa de Lima &    28.18282  &  17.22560   &  8.5          &  8.0   \\ \hline
(6) Vallehermoso  &  San Juan Bautista    &   28.18105   &   17.26570   & 15.5          & 12.5    \\ \hline
(7) La Dama, Vallehermoso  &  Ntra. Sra. de las Nieves & 28.05208 &	17.30060 & 20.5            &  10.0       \\ \hline
(8) Playa de Hermigua &	Sta. Catalina de Alejandría &	28.17927 &	17.18222 & 28.5            &   0.0      \\ \hline
(9) San Sebastián &	Ntra. Sra. de la Inmaculada Concepción &	28.08889 & 17.11480 & 36.0 &   6.0      \\ \hline
(10) Valle Gran Rey  &	Ermita de los Santos Reyes &	28.10618 &	 17.32376  &	40.0       &  14.5       \\\hline
(11) Agulo &	Ntra. Sra. de las Mercedes &	28.18890	&  17.19412	&   42.5           &   0.0      \\\hline
(12) Hermigua &	San Juan Bautista  &	28.16167 &	17.20231 &	48.5                       & B -0.5       \\\hline
(13) Hermigua &		Ntra. Sra. de la Encarnación &	 28.16863  & 17.19410 &  49.5              &   0.0      \\\hline
(14) Playa de Valle Gran Rey &		San Pedro Apóstol & 28.09509 & 17.34207 &  52.0            &  20.5       \\\hline
(15) Arure & Ntra. Sra. de la Salud  &	28.13326 & 17.31979 & 	53.5                               &   5.0      \\\hline
(16) San Sebastián &		Ermita de San Sebastián	 &	28.09379 & 17.11232 & 53.5         &  26.5       \\\hline
(17) San Sebastián &		Ntra. Sra. de la Asunción & 	28.09258  & 17.11121 & 	59.5       &  19.5       \\\hline
(18) Valle Gran Rey &		San Antonio de Padua & 	28.11781 & 	17.31298 & 	60.5       &  19.5      \\\hline
(19) Las Hayas      &		Ntra. Sra. de Coromoto  &		28.12982 &		17.29000 &  66.5    &  10.5    \\\hline
(20) Alajeró	&  Santiago Apóstol &		28.03289 &		17.19202	 &	66.5                &  17.5    \\\hline
(21) Punta Llana &		Ntra. Sra. de Guadalupe (ermita antigua) &		28.12657 & 17.10388 & 66.5  & B 2.5   \\\hline
(22) Tejiade	 &	San José	 &	28.07216 &		17.19218 &		78.5                &   1.0   \\\hline
(23) Chipude     &		Ntra. Sra. de la Candelaria &		28.10984	 &	17.28239  & 84.0    &   7.0   \\\hline
(24) Alajeró	 &	El Salvador	 &	28.06347 &		17.24038 &		88.5                &   5.0   \\\hline
(25) Roque Calvario, Alajeró	 &	San Isidro	 &	28.05301	 &	17.24199	 & 109.5    &  -1.0    \\\hline
(26) Alojera, Vallehermoso &		Ntra. Sra. de la Inmaculada Concepción	 &	28.16053 &  17.32483  &	117.5  &  20.0     \\\hline
(27) Erque, Alajeró &	San Lorenzo	 &	28.08335 &		17.26091	 &	126.0                  &  12.5  \\\hline
(28) Benchijigua    &	San Juan Bautista	 &	28.09157	 &	17.21880	 &	177.5          &  16.5  \\\hline
(29) Igualero       &	San Francisco  &		28.09940 &		17.25491 &		231.0         & B -1.5  \\\hline
(30) Agulo	 &	San Marcos Evangelista    &		28.19772 &		17.19783 &		239.5  &  13.5    \\\hline
(31) Las Nieves	 &	Ntra. Sra. de la Salud	 &	28.10112 &		17.20201 &		255.5          &   1.5   \\\hline
(32) Valle Gran Rey    &	Ntra. Sra. del Buen Viaje  &		28.13870 &		17.33758 &  285.0      &   0.0   \\\hline
(33) Playa de Santiago &	Virgen del Carmen &		28.02728 &		17.19864  & 	289.5          & B 0.0   \\\hline
(34) La Palmita        &	San Isidro &		28.17124 &		17.21595	 &	290.0          &  13.5    \\\hline
(35) Guarimiar &		Sagrado Corazón de Jesús &		28.06678 &		17.22075  & 297.0      &  23.5    \\\hline
(36) Parque Nac. de Garajonay	 &	 Ntra. Sra. de Lourdes  &		28.12696 & 	17.22085  & 302.5      & B 0.0   \\\hline
(37) Valle Gran Rey    & Virgen del Carmen                         &	28.08205 &  	17.33348  &	309.0          &   0.0   \\\hline
(38) Hermigua          & Ntra. Sra. del Rosario (conv. San Pedro)  &		28.15278  & 17.19759 & 325.5           &  13.5    \\\hline
(39) Punta Llana       & Ntra. Sra. de Guadalupe (iglesia actual)  &	28.12657 &		17.10388 & 	334.5  &   4.0   \\\hline

\end{tabular}
\end{adjustbox}
\end{table*}

\begin{figure}[ht]
\centering
\includegraphics[width=0.48\textwidth]{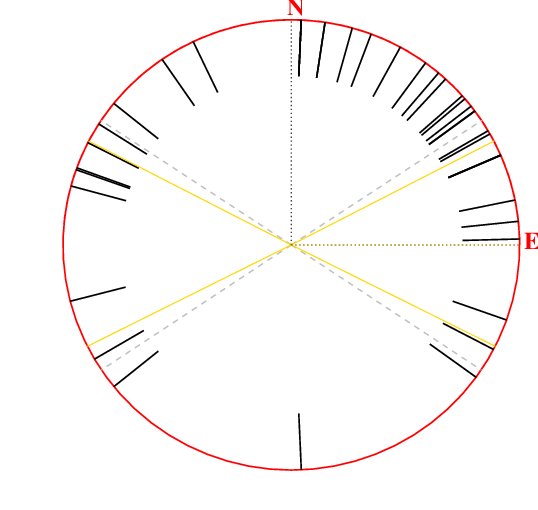}
\caption{Diagrama de orientación para las iglesias y ermitas de La Gomera, obtenido a partir de los datos de la Tabla 1. Aunque varias (unas doce) construcciones siguen la orientación canónica en el
rango solar, un gran número de iglesias está orientado hacia el noreste. Este patrón, a priori inesperado, requiere una explicación.}
\label{fig:fig-diagr-orient}
\end{figure}

En la Fig. \ref{fig:fig-diagr-orient} se muestra el diagrama de orientación para las iglesias y ermitas estudiadas. Como hemos mencionado, los valores de los acimuts son los medidos, e incluyen la
corrección por declinación magnética. Las líneas diagonales del gráfico señalan los acimuts correspondientes -en el cuadrante oriental- a los valores extremos para el Sol (acimuts de 62,8$^{\circ}$ y
116,5$^{\circ}$ -líneas continuas-, equivalente a los solsticios de verano e invierno boreales, respectivamente) y para la Luna (acimuts: 56,7$^{\circ}$ y 123,5$^{\circ}$ -líneas rayadas-, equivalente
a la posición de los lunasticios mayores).

\section{LA ORIENTACIÓN DE IGLESIAS Y EL PAISAJE}

De las 39 orientaciones medidas en las ermitas e iglesias, 13 se dirigen hacia el cuadrante norte, 9 hacia el cuadrante occidental, 16 hacia el oriental y tan sólo una hacia el cuadrante
meridional. De todas estas, sólo 12 orientaciones se ubican en el rango solar, ya sea a levante (7) o a poniente (5). Dado que nuestra muestra es representativa de toda la isla de La Gomera, estos
datos nos permiten inferir un cierto patrón de orientación. Del diagrama de la Fig. \ref{fig:fig-diagr-orient} se distingue una orientación clara, hacia el noreste, que no guarda precedente en otros
estudios de iglesias antiguas o coloniales, a excepción quizá de aquellas de Lanzarote.\cite{gangui-etal-16} Pero como veremos, en ese caso la razón es diferente.

Las 7 orientaciones {\it ad orientem} entran dentro de la lógica observada en otros estudios sobre orientaciones de iglesias, pero en nuestro caso, como dijimos, resulta notable la cantidad de
orientaciones hacia el noreste y que caen fuera del rango solar. Parece tratarse de un caso singular de esta isla, que creemos está relacionado con la orografía particular que domina el territorio,
como discutiremos brevemente más adelante.

\begin{figure}[ht]
\centering
\includegraphics[width=0.24\textwidth]{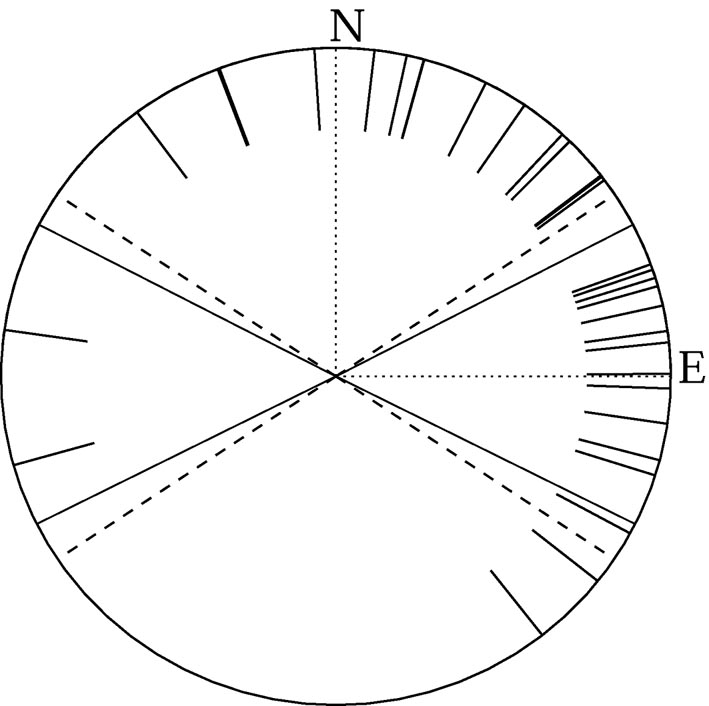}
\includegraphics[width=0.24\textwidth]{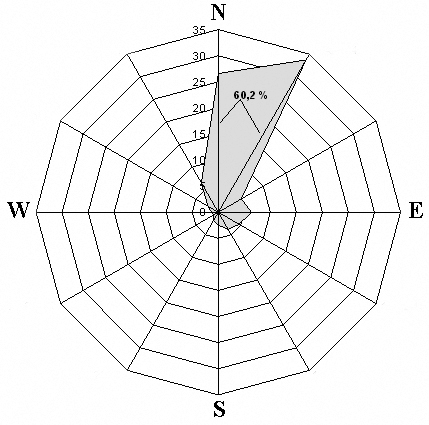}
\caption{Diagrama de orientación para las iglesias de Lanzarote (izquierda) y diagrama de vientos característico para esa misma isla (derecha), ilustrativo de los vientos alisios dominantes que se
  piensa han llevado a sus pobladores a construir las iglesias apartándose de las orientaciones ``canónicas''.\cite{gangui-etal-16}} 
\label{fig:fig-lanzarote-x2}
\end{figure}

Por el momento, recordemos que en la isla de Lanzarote (Fig. \ref{fig:fig-lanzarote-x2}) una notable proporción de las construcciones se orientaba aproximadamente hacia el norte-noreste (con
``entrada'' a sotavento) para evitar los vientos dominantes del lugar, los alisios provenientes justamente de esa dirección. Pero en La Gomera ese factor ``práctico'' lanzaroteño, debido a la
particular combinación de geografía y clima (las tormentas de arena de la región de El Jable de esa isla) no es preponderante.

La particularidad en las orientaciones de las iglesias gomeras tiene poco correlato con otros estudios previos. En la Península Ibérica y en todo el Mediterráneo los rangos son predominantemente
solares. En especial, la gran proporción de iglesias orientadas aproximadamente hacia el noreste resulta novedoso. Este es el caso de la iglesia Nuestra Señora de la Encarnación, y de las ermitas San
Juan Bautista y Santa Catalina, todas ellas ubicadas a lo largo del barranco (o valle) de Hermigua. Lo mismo sucede con las iglesias San Antonio de Padua y San Pedro Apóstol, y con la ermita de los
Santos Reyes, también ubicadas a lo largo de un barranco prominente, pero que esta vez se trata del Valle Gran Rey, hacia la costa oeste de la isla (ver Tabla 1).

La presencia de amplios y profundos barrancos, generados en el suelo volcánico de la isla por la erosión del agua, es una de las marcas características del paisaje de La Gomera. Basándonos en este
trabajo arqueoastronómico, aún en progreso, creemos que ese mismo paisaje es el que marcó la particular orientación de una buena parte del conjunto de sus iglesias históricas. 

\begin{figure*}[ht]
\centering
\includegraphics[width=0.82\textwidth]{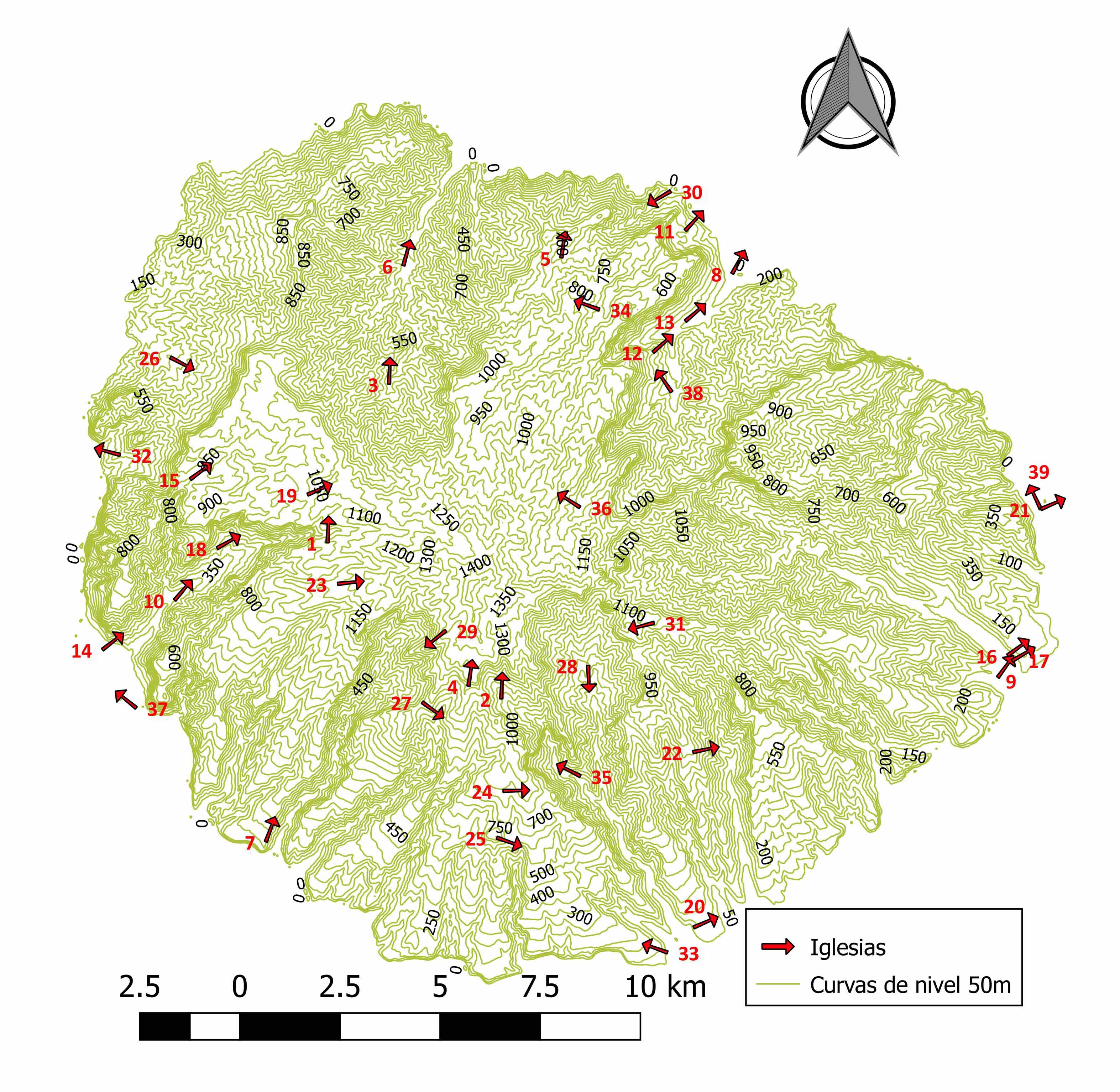}
\caption{Mapa topográfico con la ubicación geográfica de la totalidad de las iglesias medidas, junto con la orientación de sus ejes en dirección al altar (flechas, orientadas de acuerdo a los
  acimuts consignados en la Tabla 1). En la región de Punta Llana (costa este) la iglesia Nuestra Señora de Guadalupe presenta dos flechas con orientaciones ortogonales. Estas se deben a que, adosada a
  la antigua ermita (núm 21), que inicialmente apuntaba con acimut 66,5$^{\circ}$ en dirección al volcán Teide, en la vecina isla de Tenerife, tiempo más tarde se construyó la nave principal de la
  iglesia actual (núm 39), con una orientación aproximadamente perpendicular.} 
\label{fig:fig-mapa-curvas-flechas}
\end{figure*}

Para visualizar mejor la posible influencia sobre las orientaciones de la orografía, en forma de accidentes naturales como valles y barrancos, o incluso por la presencia de algún monte prominente de
la isla (como el Roque de Imada) o de islas cercanas (como el volcán Teide de la isla de Tenerife, bien visible desde La Gomera), en la Fig. \ref{fig:fig-mapa-curvas-flechas} presentamos un mapa
topográfico que incorpora las orientaciones de todas las iglesias estudiadas.\footnote{El mapa fue realizado a partir del procesamiento de imágenes LANDSAT, cortesía de USGS (Servicio geológico de los
  Estados Unidos).}

En este mapa se aprecia que los dos grupos de tres iglesias ya mencionados ``copian'' la dirección de los valles profundos en los que se hallan emplazados, y en ambos casos los barrancos siguen una
línea sudoeste-noreste coincidente con la acumulación de orientaciones en la región noreste de nuestra Fig. \ref{fig:fig-diagr-orient}. A esto se suma la presencia en la propia capital de la isla, San
Sebastián de La Gomera, de tres construcciones con orientaciones en el cuadrante nor-oriental: las ermitas Nuestra Señora de la Inmaculada Concepción y de San Sebastián, y la iglesia matriz Nuestra
Señora de la Asunción (ver datos en Tabla 1) que, con buena aproximación, refuerzan ese patrón.

La orografía también permite comprender la disposición de otras construcciones que podríamos llamar ``anómalas'', en el sentido de que se apartan mucho de las esperadas, si nos atenemos a la tradición
religiosa canónica de orientaciones. Ejemplo de esto es la iglesia Santa Ana, en Imada, que se orienta prácticamente con la meridiana del lugar (con un acimut de sólo 2,5$^{\circ}$, en dirección
norte). Una inspección del paisaje circundante, sin embargo, muestra que el eje de la iglesia se alinea con muy buena aproximación con la elevación montañosa más distintiva del lugar: el ya mencionado
Roque de Imada (la palabra ``roque'', empleada en Canarias, denota una elevación rocosa y muy escarpada). Pero en este caso, el roque se halla del lado de la puerta de la iglesia, en dirección opuesta
a la del altar, por lo que su visión se vuelve imponente al salir del edificio, luego de culminar el oficio religioso. Guardando las distancias -tanto geográficas como por tratarse de diferentes
regiones culturales-, casos similares a estos ya fueron descritos y estudiados con cierto detalle en el caso de las iglesias coloniales andinas del norte de Chile.\cite{gangui-16} 

Otro ejemplo es el de la ermita Nuestra Señora del Buen Paso, en Alajeró, cuyo eje tiene un acimut de 8,5$^{\circ}$ y, por lo tanto, se aparta mucho de las orientaciones canónicas. Nuevamente en este
caso, la inspección del paisaje circundante nos indica que el movimiento del Sol, y el arco de sitios del horizonte por donde surge o se oculta en diferentes días del año, fue irrelevante para quienes
concibieron la iglesia. Pues la construcción se ubica pegada a la ladera de un barranco escarpado, con el fin presumible de indicar el mejor camino para atravesar la montaña, y por supuesto no hubo
libertad para orientarla adecuadamente.

\section{DISCUSIÓN Y TRABAJO FUTURO}

Luego de la ocupación de la zona de San Sebastián -futura capital de la isla- por parte de Hernán Peraza ``el Viejo'', caballero al servicio de la corona de Castilla, en las décadas siguientes del
siglo XV se inició la colonización más amplia del territorio insular con el establecimiento de haciendas y caseríos.\cite{darias-92} En muchas de estas villas, el crecimiento de la población fue
acompañado por la construcción de pequeñas ermitas y templos cristianos que ilustraban la nueva situación religiosa y social.

En algunos lugares, es posible que los templos cristianos se orientasen con patrones de imitación del culto aborigen, prehispánico, especialmente en direcciones solsticiales.\cite{belmonte-etal-94}
Tal podría ser el caso de la iglesia Nuestra Señora de la Inmaculada Concepción, de Alojera, cuyo eje se orienta con un acimut de 117,5$^{\circ}$, a sólo un grado de la dirección del Sol naciente
durante el solsticio de invierno boreal. Sin embargo, sabemos que el orto solar en fechas cercanas al solsticio de invierno fue una orientación muy poco usada en el mundo ibérico cristiano. Además, el
paisaje circundante a esta iglesia, que se cree data de mediados del siglo XVII, muestra un horizonte montañoso, sobre todo en dirección hacia el altar (con $h=20^{\circ}$). 
Sabemos que un análisis completo requiere tomar 
en cuenta la altura angular de dicho horizonte, pues un perfil elevado detrás de la iglesia cambiará la fecha en la que el Sol en el horizonte podría alinearse con su eje.

Este estudio que señalamos para La Concepción de Alojera, también deberá implementarse para toda la muestra de las iglesias de la isla. Un trabajo futuro, aún en progreso, nos permitirá combinar
medidas locales de acimut y altura angular para obtener la declinación, coordenada ecuatorial que ya no dependerá de la ubicación geográfica ni de la topografía local. El valor de esta coordenada,
estimado para una dada iglesia, una vez comparado con la declinación del Sol (que fija aproximadamente un par de días en el año, o sólo uno en el caso de los solsticios), nos permitirá verificar,
entre otras cosas, si esa iglesia está o no orientada en una dirección que coincide con la fecha de su fiesta patronal,
y evaluar el peso estadístico de estos resultados. 

Por otra parte, en ciertos lugares de la isla se respetó la tradición canónica de orientar los templos a levante, aunque con un grado de libertad claramente mayor que el habitual. Ejemplo de esto son
dos de las iglesias más antiguas y emblemáticas: Nuestra Señora de la Candelaria, en Chipude (construida en la segunda mitad del siglo XVII), y El Salvador, en Alajeró (de aproximadamente el año
1666), que están orientadas de manera aproximadamente equinoccial.

A diferencia de las anteriores, la iglesia Nuestra Señora de Guadalupe, en Punta Llana, es curiosa por la coexistencia de una doble orientación, en la que la ermita antigua se orienta con un acimut de
66,5$^{\circ}$ y la nave moderna actual se dispone de manera aproximadamente perpendicular (Fig. \ref{fig:fig-mapa-curvas-flechas}). La pequeña ermita original, hoy con función de capilla lateral de
la iglesia principal, ubicada a la derecha del altar, posee una orientación casi solsticial y, además, apunta directamente hacia el volcán Teide, la elevación más prominente de la vecina isla de
Tenerife. Como discutimos antes, este es un caso en donde la orientación solar y la orografía insular compiten en plano de igualdad. Más datos y nuevos análisis, actualmente en curso, nos permitirán
dirimir la cuestión.

Del resto de las iglesias, como ya mencionamos, creemos que una gran proporción se adapta a la geografía de sus sitios de emplazamiento, y se orienta de acuerdo a los barrancos, o incluso cabeceras de
playa, en donde se fueron ubicando. A partir del estudio que hemos presentado podemos conjeturar que, si bien los primeros pobladores intentaron mantener las orientaciones canónicas en las iglesias que
iban construyendo, en relativamente poco tiempo se dieron cuenta de que el territorio les jugaba en contra. Esto no es extraño, dada la ``abrupta naturaleza'' de La Gomera.\cite{diaz-05} 

Muchas de estas iglesias, en particular aquellas de los barrancos de Hermigua y de Valle Gran Rey, parecen ser las responsables de la acumulación de orientaciones en la región noreste de la
Fig. \ref{fig:fig-diagr-orient}. Sin embargo, el análisis completo deberá incluir otros datos, aún en estudio, como ser el perfil montañoso circundante a los templos religiosos. Este análisis 
estadístico 
nos proveerá la distribución del número aproximado de iglesias por cada valor de declinación posible. Con ese dato podremos entonces verificar fehacientemente si la acumulación de orientaciones en el
diagrama de acimuts deja su marca también en un gráfico de declinaciones, que es, en fin de cuentas, aquello que nos señala la posible influencia astronómica -más especificamente, del movimiento del
Sol- en la orientación de las iglesias históricas.

\subsection{Agradecimientos}

Los autores desean agradecer a sus colaboradores Juan Belmonte, A. César González-García y María A. Perera-Betancort, por frecuentes discusiones en estos temas y por su constante ayuda y apoyo.
Agradecen también los comentarios críticos de un evaluador anónimo que sirvieron para aclarar algunos puntos de la presentación.
El trabajo de A.G. ha sido financiado parcialmente por CONICET y por la Universidad de Buenos Aires.


\section{REFERENCIAS
  \vspace{-3ex}}  

\bibliographystyle{unsrt}

\begin{thebibliography}{1}

\bibitem{gonzalez-15}
A~González-García.
\newblock A {Voyage} of {Christian} {Medieval} {Astronomy}: Symbolic, ritual
  and political orientation of churches.
\newblock In {\em Stars and Stones: Voyages in archaeoastronomy and cultural
  astronomy}. British Archaeology Reports, Int. Ser. 2720, {edited by F.
  Pimenta et al.}, 268-275, 2015.

\bibitem{mccluskey-15}
S~McCluskey.
\newblock Orientation of christian churches.
\newblock In C~Ruggles, editor, {\em Handbook of Archaeoastronomy and
  Ethnoastronomy}. New York, Springer-Verlag, 1703-1710, 2015.

\bibitem{zimbron-15}
J~R Zimbrón~Romero and R~Moyano.
\newblock La fiesta de la {Virgen} de {Guadalupe} asociado a un marcador
  pre-solsticial en la parte norte de la {Cuenca} de {México}.
\newblock In {\em Diferentes povos, diferentes saberes na América Latina.
  Contribuções da astronomia cultural para a história da ciência. Actas del
  congreso SIAC 2013}. Museu de Astronomia e Ciências Afins, {edited by L.C.
  Borges}, 126-151, 2015.

\bibitem{gangui-etal-16}
A~Gangui, A~González-García, M~Perera-Betancort, and J~Belmonte.
\newblock La orientación como una seña de identidad cultural: las iglesias
  históricas de {Lanzarote}.
\newblock {\em Tabona: Revista de Prehistoria y Arqueología}, 20:105--128,
  2016.

\bibitem{diaz-05}
G~Díaz~Padilla.
\newblock La evolución parroquial de {La Gomera} y el patrimonio documental
  generado por la institución eclesiástica.
\newblock {\em Memoria ecclesiae}, Vol. 27:365--376, 2005.

\bibitem{vogel-62}
C~Vogel.
\newblock Sol aequinoctialis. {Problèmes} et technique de l'orientation dans
  le culte chrétien.
\newblock {\em Revue des Sciences Religieuses}, Vol. 36:175--211, 1962.

\bibitem{belmonte-etal-94}
J~Belmonte, C~Esteban, A~Aparicio, A~Tejera~Gaspar, and O~Gónzalez.
\newblock Canarian {Astronomy} before the conquest: the pre-hispanic calendar.
\newblock {\em Rev. Acad. Can. Ciencias}, Vol. VI(2-3-4):133--156, 1994.

\bibitem{gangui-16}
A~Gangui.
\newblock Archaeoastronomy and the orientation of old churches.
\newblock {\em Bulletin of the Argentine Astronomical Society (BAAA)}, Vol.,
  58:paper \#261, P. Benaglia, et al. (eds.) 2016.

\bibitem{darias-92}
A~Darías~Príncipe.
\newblock {\em La Gomera: espacio, tiempo y forma}.
\newblock Compañía Mercantil Hispano-Noruega, Santa Cruz de Tenerife, 1992.

\end{thebibliography}
\renewcommand{\refname}{}

\end{document}